\documentclass[12pt]{article}
\usepackage{amsmath}
\usepackage{graphicx,psfrag,epsf}
\usepackage{enumerate}
\usepackage{natbib}
\usepackage{textcomp}
\usepackage[hyphens]{url} 
\usepackage{hyperref}

\newcommand{\blind}{0}

\providecommand{\tightlist}{%
  \setlength{\itemsep}{0pt}\setlength{\parskip}{0pt}}

\usepackage{xcolor}
\xdefinecolor{blue}{rgb}{0.28,0.65,0.9}
\hypersetup{
    colorlinks,
    linkcolor={blue},
    citecolor={blue},
    urlcolor={blue}
}

\usepackage{booktabs}

\newcommand{\Duke}[1]{Duke University}
\newcommand{\Muser}[1]{Muser}
\newcommand{\Course}[1]{STA 210: Regression Analysis}
\newcommand{\DukeBlind}[1]{a four-year institution}
\newcommand{\MuserBlind}[1]{Online Platform}
\newcommand{\CourseBlind}[1]{Regression Analysis}

\usepackage{booktabs}
\usepackage{longtable}
\usepackage{array}
\usepackage{multirow}
\usepackage{wrapfig}
\usepackage{float}
\usepackage{colortbl}
\usepackage{pdflscape}
\usepackage{tabu}
\usepackage{threeparttable}
\usepackage{threeparttablex}
\usepackage[normalem]{ulem}
\usepackage{makecell}
\usepackage{xcolor}

\begin{document}

\def\spacingset#1{\renewcommand{\baselinestretch}%
{#1}\small\normalsize} \spacingset{1}


\if0\blind
{
  \title{\bf Three principles for modernizing an undergraduate
regression analysis course}

  \author{
        Maria Tackett \\
    Department of Statistical Science - Duke University\\
      }
  \maketitle
} \fi

\if1\blind
{
  \bigskip
  \bigskip
  \bigskip
  \begin{center}
    {\LARGE\bf Three principles for modernizing an undergraduate
regression analysis course}
  \end{center}
  \medskip
} \fi

\bigskip
\begin{abstract}
As data have become more prevalent in academia, industry, and daily
life, it is imperative that undergraduate students are equipped with the
skills needed to analyze data in the modern environment. In recent years
there has been a lot of work innovating introductory statistics courses
and developing introductory data science courses; however, there has
been less work beyond the first course. This paper describes innovations
to Regression Analysis taught at \Duke{}, a course focused on
application that serves a diverse undergraduate student population of
statistics and data science majors along with non-majors. Three
principles guiding the modernization of the course are presented with
details about how these principles align with the necessary skills of
practice outlined in recent statistics and data science curriculum
guidelines. The paper includes pedagogical strategies, motivated by the
innovations in introductory courses, that make it feasible to implement
skills for the practice of modern statistics and data science alongside
fundamental statistical concepts. The paper concludes with the impact of
these changes, challenges, and next steps for the course. Portions of
in-class activities and assignments are included in the paper, with full
sample assignments and resources for finding data in the supplemental
materials.
\end{abstract}

\noindent%
\vfill

\newpage
\spacingset{1.45} 

\begin{center}\rule{0.5\linewidth}{0.5pt}\end{center}

\hypertarget{introduction}{%
\section{Introduction}\label{introduction}}

In recent years there has been a lot of innovation in introductory
statistics courses; however, there has been less work on innovating
subsequent courses for the modern student and data environment. It is
important that courses beyond the first one are modernized as well, so
students can continue developing the necessary skills to analyze data
and effectively communicate results as they progress through the
statistics curriculum.

This paper focuses on modernizing the undergraduate regression analysis
course at \Duke{} aimed at a broad audience of quantitatively-minded
students from majors across the university. As the Statistical Science
major has grown substantially and as more students from across the
university conduct more data-driven work in and outside of the
classroom, it is imperative to offer a second course that can prepare
non-majors to use statistical methods in research and internships while
also preparing majors for the statistical and computational rigor of
upper-level courses. The 2018 report by the National Academies of
Sciences, Engineering, and Medicine emphasizes the imperative to equip
all undergraduate students with such skills in the recommendation that
``academic institutions should encourage the development of a basic
understanding of data science in all undergraduates.''
\citep{national2018data}. This understanding includes many of the
knowledge and skills in the second course, such as statistical modeling,
model assessment, and model interpretation, among others.

Taking into account the skills students need in the modern data-driven
environment, three principles have motivated the innovations to the
pedagogy in the regression analysis course at \Duke{}.

\begin{quote}
\emph{Facilitate the opportunity for students to}

\emph{1. Regularly engage with complex {[}and relevant{]} real-world
data and applications}
\end{quote}

\begin{quote}
\emph{2. Develop the skills and computational proficiency for a
reproducible data analysis workflow}
\end{quote}

\begin{quote}
\emph{3. Develop important non-technical skills, specifically written
communication and teamwork}
\end{quote}

These principles are largely inspired by innovations in introductory
statistics courses and the creation of introductory data science
courses. They are also driven by the skills students need as they apply
for graduate school, internships, and careers. For example, a recent ad
for the position \emph{Staff Editor - Statistical Modeling} at the
\emph{New York Times} asked applicants to ``describe or link to an
example of a statistical model you've created\ldots describe any
reporting, development or data visualization skills you may have.''
\citep{nyt}. The ad also listed expertise with R and statistical
modeling as requirements for the position.

The remainder of the paper focuses on the implementation of these three
principles in an undergraduate regression analysis course at \Duke{}.
Section \ref{background} provides a summary of similar second courses at
other institutions and a brief overview of statistics guidelines for
instruction. The course structure and pedagogy are detailed in Section
\ref{course-description}, and Section \ref{principles} expands on the
guiding principles used for modernizing the courses and the pedagogy of
implementing these changes. Section \ref{discussion} concludes with a
discussion of the impact of the changes, challenges, and next steps for
the course.

\hypertarget{background}{%
\section{\texorpdfstring{Background
\label{background}}{Background }}\label{background}}

\hypertarget{the-course-in-the-statistics-curriculum}{%
\subsection{The course in the statistics
curriculum}\label{the-course-in-the-statistics-curriculum}}

\Course{} is the second statistics course for many students who have
taken an introductory statistics, data science, or probability course at
\Duke{}. Taking one of these introductory courses is a prerequisite
requirement for the class, so most students have some prior experience
conducting exploratory data analysis, statistical inference, and simple
linear regression using R. Students who have taken the introduction to
data science course also have prior experience writing reproducible
reports using R Markdown and implementing version control using git and
GitHub. Students who take probability as a first course generally have
less prior experience with data analysis and computing; however, they
have more in-depth prior study of the mathematics of probability and
statistics.

This regression course has a few important roles in the undergraduate
statistics curriculum. It is a core requirement of the undergraduate
Statistical Science major and minor, and a pre-requisite for most of the
upper-level courses in the Department of Statistical Science. It is also
the earliest course taken by all students in the major and minor, so it
is their first shared experience in the statistics curriculum.
Therefore, it is one of the first opportunities to teach fundamental
skills that are part of the learning outcomes for the statistics
undergraduate curriculum \citep{statsci-LOs}, in particular

\begin{itemize}
\tightlist
\item
  \emph{Students will demonstrate ability in computational
  methods---including basic statistical programming, data analysis, and
  reproducibility---necessary to do applied data analysis.}
\item
  \emph{Students will demonstrate the ability to use appropriate
  statistical methodologies for real-world data analysis settings.}
\item
  \emph{Students who take only one or two courses from the department
  will demonstrate understanding of the usefulness, importance, and
  power of statistical thinking and methodologies.}
\end{itemize}

In addition to the course's role in the statistics curriculum, it is
also a service course that is taken by students with a variety of
academic interests. The student population includes students who are or
intend to be Statistical Science majors and students in majors from
other disciplines who are interested in developing their analysis skills
in preparation for data-driven research, graduate school, or
internships. Thus the course content, learning goals, and instructional
design are developed to serve the diverse student population with
competing learning objectives.

\hypertarget{related-courses-at-other-institutions}{%
\subsection{Related courses at other
institutions}\label{related-courses-at-other-institutions}}

This paper is about innovations to the undergraduate regression analysis
course taught at \Duke{}; however, there are undergraduate regression
analysis courses at other institutions that are similarly focused on
application and are implementing some of the pedagogical approaches
described in the text. A few of these courses are

\begin{itemize}
\tightlist
\item
  STA 212: Statistical Models at Wake Forest University
\item
  STAT 272: Statistical Modeling at St.~Olaf College
\item
  STA 363: Intro to Statistical Modeling at Miami University
\item
  STA 324: Applied Regression Analysis at California Polytechnic State
  University
\end{itemize}

Based on a review of available course descriptions, syllabi, and
websites, these courses are similar to the regression analysis course
discussed in the paper in that most occur earlier in the statistics
curriculum and generally emphasize application over mathematical theory.
Focusing on application helps better serve both non-majors and majors,
as it prepare students who may not take additional statistics courses
while also motivating the theory majors encounter in advanced courses.

\hypertarget{curriculum-guidelines-and-related-work}{%
\subsection{Curriculum guidelines and related
work}\label{curriculum-guidelines-and-related-work}}

The \emph{Guidelines for Assessment and Instruction in Statistics
Education (GAISE)} \emph{College Report} \citep{carver2016guidelines}
gives guidance on the skills students should develop in their
introductory course and pedagogy recommendations to help students
achieve these aims. Though the report is specifically focused on the
introductory course, the authors of the report state that their
recommendations extend beyond the introductory course and could be
applied throughout the undergraduate curriculum. They encourage
statistics instructors to ``emphasize the practical problem-solving
skills that are necessary to answer statistical questions.'' (pg. 12)
They also emphasize the use of real-world data with context in classes,
as using ``real data sets of interest to students is a good way to
engage students in thinking about the data and relevant statistical
concepts.'' (pg. 17), and as part of that, exposing students to the
messiness that often arises when working with real data.

The American Statistical Association's 2014 \emph{Curriculum Guidelines
for Undergraduate Programs in Statistical Science}
\citep{american2014curriculum} provides recommendations about the skills
that are important for an undergraduate statistics major. In addition to
the knowledge of statistical methods and theory, the report emphasizes
skills in statistical practice that are critical for students as they
prepare for careers in statistics and data science. Some of the skills
it highlights are key drivers for the principles guiding the
modernization of the regression course. The report states that courses
should focus on the use of ``authentic data'' and that the curriculum
should include ``concepts and approaches for working with complex
data\ldots and analyzing non-textbook data'' (Principle 1).
Additionally, the report states that students should be ``facile with
professional statistical software'' and that students' analyses ``should
be undertaken in a well-documented and reproducible way'' (Principle 2).
Finally, the report discusses the importance of teaching students skills
in ``statistical practice'' including being able to ``write clearly,
speak fluently, and construct effective visual displays and compelling
written summaries'' and ``demonstrate ability to collaborate in teams
and to organize and manage projects'' (Principle 3).

While there is no expectation students would demonstrate mastery of
these skills by the end of the second course, emphasizing them earlier
in the curriculum allows time to better equip statistics majors with the
full suite of skills needed to use statistics and data science in
practice. Much of this philosophy has been incorporated in the
introductory courses. In recent years there has been a wealth of
literature on innovations in introductory statistics and data science
courses
\citep[e.g.,][]{adams2021computational, ccetinkaya2021fresh, farmus2020flipped, baumer2015data, hardin2015data}.
Many of these innovations, have been motivated by the need to help
students gain the conceptual knowledge and computing skills required to
analyze authentic complex and nonstandard data, such as analyzing text
and spatial data. Additionally these newly revised courses have put more
emphasis on visualizing and interpreting multivariable relationships
\citep{adams2021computational} in line with \emph{GAISE} recommendation
to ``give students experience with multivariable thinking.'' (pg. 6)

Despite the abundance of literature on introductory statistics and data
science courses, there has been less published work about subsequent
undergraduate courses, especially the ``second'' statistics course.
\citet{love1998project} and \citet{roback2003teaching} describe courses
primarily focused on linear and logistic regression aimed at a broad
student audience, with \citet{love1998project} presenting a
project-based approach for teaching this content.
\citet{blades2015second} makes the point that even though linear
regression has been taught in most second statistics courses, given the
relative recent development of many of these courses there isn't a
consensus about the content that should be covered. In fact they propose
focusing on the design and analysis of experiments rather than
regression analysis as the next course in statistics.

Taking the curriculum guidelines and previous work into account, much of
the innovation in the course has been in regard to skills important for
modern data analysis and pedagogical strategies for teaching these
skills. As seen in the next section, many of the statistical topics
align with those more traditionally taught in an undergraduate
regression or second statistics course.

\hypertarget{pedagogy}{%
\section{\texorpdfstring{Pedagogy
\label{course-description}}{Pedagogy }}\label{pedagogy}}

\hypertarget{learning-objectives-and-topics}{%
\subsection{Learning objectives and
topics}\label{learning-objectives-and-topics}}

Below are the primary learning objectives for the course:

\begin{quote}
\emph{By the end of the semester students will be able to\ldots{}}

\begin{itemize}
\tightlist
\item
  \emph{analyze real-world data to answer questions about multivariable
  relationships.}
\item
  \emph{fit and evaluate linear and logistic regression models.}
\item
  \emph{assess whether a proposed model is appropriate and describe its
  limitations.}
\item
  \emph{use R Markdown to write reproducible reports and GitHub for
  version control and collaboration.}
\item
  \emph{communicate results from statistical analyses to a general
  audience.}
\end{itemize}
\end{quote}

The course is divided into three units: linear regression, logistic
regression, looking ahead.

\textbf{Linear regression (Weeks 01 - 09)}: The unit includes a review
of statistical inference, simple and multiple linear regression, and
ANOVA. Topics include the interpretation of the model coefficients,
inference for coefficients and predictions, assessing model conditions
and diagnostics, categorical predictors, interactions and polynomial
predictors, log-transformations on the response and predictor variables,
and model selection.

\textbf{Logistic regression (Weeks 10 - 12)} :The unit primarily focuses
on logistic regression with a brief introduction to multinomial logistic
regression. Topics include the interpretation and inference of model
coefficients, model conditions, selection, and using the ROC curve to
assess model fit and prediction.

\textbf{Looking ahead (Weeks 13 - 15)}: The unit is a collection of
special topics that may vary each semester. The purpose of the special
topics is to introduce students to models and related methods that
extend beyond the scope of the course. Recent topics have included
dealing with missing data, models for correlated data, time series, and
model validation. There is also one lecture in the last week of classes
dedicated to advanced skills to write reports using R Markdown (e.g.,
including citations, figure captions, etc.).

A more detailed outline of the course topics and schedule is available
in the supplemental materials in Section \ref{supplement}

\hypertarget{week-schedule}{%
\subsection{Course structure and assessments}\label{week-schedule}}

About 90 - 100 students take the course each semester, and students
attend two full-class 75-minute lectures and one smaller 75-minute lab
each week. The lectures are primarily focused on the introduction of new
statistical concepts, and the labs are focused on computing and
application. Before each lecture, students complete a \emph{prepare}
assignment that includes a combination of videos and readings. The
prepare assignments introduce students to new definitions, concepts and
brief mathematical details, and an example demonstrating how the concept
is applied. Students have access to the slides presented in the videos,
so the prepare assignment primarily replaces the presentation of slides
that is common in more traditional lecture formats. The lecture sessions
follow a flipped format with hands-on exercises and applications. The
flipped lectures are detailed in Section \ref{flipped-lecture}. A
typical week in the course is outlined in Table \ref{tab:schedule}.

\begin{longtable}[t]{>{\raggedright\arraybackslash}p{4cm}|>{\raggedright\arraybackslash}p{6cm}}
\caption{\label{tab:schedule}Example week in the course}\\
\toprule
Day & Activity\\
\midrule
\cellcolor{gray!6}{Monday} & \cellcolor{gray!6}{Complete prepare assignment}\\
Tuesday & In-class lecture\\
\cellcolor{gray!6}{Wednesday} & \cellcolor{gray!6}{Complete prepare assignment}\\
 & Tuesday's in-class exercises due\\
\cellcolor{gray!6}{Thursday} & \cellcolor{gray!6}{In-class lecture}\\
\addlinespace
Friday & In-class lab\\
\cellcolor{gray!6}{} & \cellcolor{gray!6}{Thursday's in-class exercises due}\\
\bottomrule
\end{longtable}

In a given semester there are usually three or four lab sections, with
25 - 30 students in each. Each lab is led by a graduate teaching
assistant with another graduate or undergraduate teaching assistant to
help answer questions. During lab sessions students work primarily in
teams of three or four on case studies involving real-world data and
analysis questions. These lab assignments account for 15\% of the final
course grade. The structure and assessment of teamwork is detailed in
Section \ref{principle3}.

In addition to weekly labs, students are assessed through individual
homework assignments, assigned about every three weeks, that account for
30\% of the final course grade. There are five assignments, four that
directly assess the content and one for a ``statistics experience''.
Students have a week to complete the four content-based assignments,
which are used to assess their ability to combine the statistical
concepts and computing skills by completing short conceptual exercises
and open-ended data analyses.

The \emph{statistics experience} is introduced early in the semester and
is due the last week of classes. The purpose of the assignment is for
students to intentionally engage with statistics outside of the
classroom by attending a talk, interviewing a statistician, listening to
relevant podcast, reading a relevant book, or participating in a data
analysis competition. They submit a slide briefly summarizing the
experience and discussing how it connects to the course content. Given
the size of the course, students submit a PDF of the slides for grading;
however, in a smaller setting it may be valuable to have students
present their experiences to the class.

The larger summative assessments in the course are periodic quizzes
(three or four per semester; 30\% of the course grade) that assess
conceptual understanding and proficiency in applying the methods on
real-world data, and a final group project (15\% of the course grade).

Over the past few years different textbooks have been used for the
course in an effort to identify an applied regression analysis text with
rich data examples that is not cost prohibitive for students. Texts such
as \emph{Stat2: Modeling and Regression with ANOVA} \citep{stat2book}
and \emph{The Statistical Sleuth: A Course in Methods of Data Analysis}
\citep{ramsey2012statistical} have been used in the course in previous
semesters. In the most recent semesters readings have come from
\emph{Introduction to Modern Statistics} \citep{imsbook} and
\emph{Handbook of Regression Analysis} \citep{chatterjee2013handbook}.

\hypertarget{flipped-lectures}{%
\subsection{\texorpdfstring{``Flipped'' lectures
\label{flipped-lecture}}{``Flipped'' lectures }}\label{flipped-lectures}}

The lectures follow a modified flipped classroom structure. Though they
do not follow the full structure in \citet{bergmann2012flip}, they are
``flipped'' in term of what students do synchronously during class
versus what they do asynchronously outside of class. This format is best
described by \citet{farmus2020flipped} as students engaging in passive
content learning prior to class, allowing the in-class time to be
dedicated to active learning of content.

Each class begins with a 5 - 10 minute review of the material from the
prepare assignment, then the rest of the lecture session is used for a
combination of individual and small group exercises where students apply
the new concepts. Through these exercises, students engage with
conceptual ideas, computational skills, interpretations, and inference,
and how these components combine to answer an analysis question. There
are multiple ways for students to participate in a given class session,
as students can work with peers, submit responses and questions using
the course discussion forum, and participate in large class discussions.
The students submit their responses to the in-class exercises at the end
of class through the GitHub repository assigned to them for that lecture
(more about the use of GitHub repositories in Section \ref{principle2}).
The responses are graded for completion, with the grading merely serving
as a measure that student are keeping on pace with the course content.
The completion grade was particularly important during remote learning
in 2020 - 2021 when some students participated fully asynchronously.

\hypertarget{shaping-the-course-to-prepare-the-modern-student}{%
\section{\texorpdfstring{Shaping the course to prepare the modern
student
\label{principles}}{Shaping the course to prepare the modern student }}\label{shaping-the-course-to-prepare-the-modern-student}}

As described thus far, the statistical topics in the course have been
similar to those often covered in a traditional regression analysis or
second statistics course, so much of the most recent innovation has been
in the pedagogical approach and content beyond the regression concepts.
The goal is to prepare students with a full suite of skills that
includes the necessary statistical knowledge, computing proficiency
required to implement the statistical methods, experience gained from
working with a variety of complex real-world data sets, skills in
effectively communicating results, and experience collaborating on data
analysis projects. As demonstrated by the work in modernizing
introductory courses discussed in Section \ref{background}, these skills
can be successfully implemented in a course without distracting from the
fundamental statistical content. Implementation has been done by
considering ways to integrate these other skills into the curriculum
alongside the statistical content, providing opportunities for practice
through in-class activities and assignments, and focusing on application
rather than mathematical theory. A half-credit course on the mathematics
of regression has been implemented in the statistics curriculum;
students have the option to take the course alongside or after
completing the regression analysis course.

Though some of these skills are introduced in the introductory course,
we seek to emphasize them in this second course for a few key reasons.
The first reason, as stated earlier, is this is the first shared course
experience among all students in the major and minor. Next, though
students are introduced to the data analysis workflow, collaboration,
and written communication in the introductory course, focusing too much
on these skills could result in cognitive overload as students are also
being introduced to statistical thinking and many new concepts. Students
come into the second course with some exposure and experience with these
skills and concepts, so there is less cognitive load as they continue
honing the skills related to workflow and collaboration alongside
learning the new statistical content. Finally, it is important that the
skills students start building in the introductory course continue to be
reinforced and further developed as they progress through the statistics
curriculum.

\hypertarget{principle-1-regularly-engage-with-complex-and-relevant-real-world-data-and-applications}{%
\subsection{\texorpdfstring{Principle 1: Regularly engage with complex
{[}and relevant{]} real-world data and applications
\label{principle1}}{Principle 1: Regularly engage with complex {[}and relevant{]} real-world data and applications }}\label{principle-1-regularly-engage-with-complex-and-relevant-real-world-data-and-applications}}

\emph{GAISE} emphasizes the importance of using real-world data in the
classroom. Though many textbooks use relevant real-world data in their
examples and exercises, the data sets often do not represent the
messiness and complexity of modern data. Therefore the ``real-world''
data referred to here are the messy data that often require cleaning and
other pre-processing before they are suitable for analysis. In previous
semesters, some students had difficulty dealing with the data cleaning
required for the final project where they analyzed a data set of their
choice. With this in mind, the data sets used in the regression course
are chosen to give students some exposure to the data cleaning and
wrangling required before doing most regression analyses, and to
demonstrate how regression can be used in a wide variety of interesting
and relevant contexts.

Working with complex and messy data sets provides some continuity as
students progress through the curriculum, as students are continuing to
hone the data wrangling skills they've learned in their introductory
courses. They are learning how to use visualizations and descriptive
statistics to explore data and make data preparation decisions, such as
how to handle missing data and outliers and the implications their
decisions have on the scope of their conclusions or potential biases in
the results. In the past they may have only dealt with issues in the
final project. While there are components of working with data that are
based on content knowledge, there are other skills that are derived from
experience and being exposed to a lot of different data sets and
contexts. By having students engage with complex data throughout the
semester, they are getting that exposure and continually honing these
skills, better preparing them as they address these questions as they
work with data outside of the classroom.

Complex real-world data sets are used in every component of the courses:
lecture slides, in-class activities, and lab and homework assignments.
Depending on the context, students will engage with more or less of the
data cleaning and pre-processing. For example, the in-class activities
mostly use cleaned data sets so that the class time can be focused on
the new regression concepts. Sometimes, however, there will be a short,
guided data cleaning exercise incorporated into the in-class activity so
we can discuss strategies to approach such tasks as a class.

One aspect that has been emphasized through the use of these data are
the importance of exploratory data analysis and how visualizations can
help inform the data cleaning and preparation process. Examples of how
this idea has been implemented in an in-class activity and in a homework
assignment are below.

\begin{figure}

{\centering \includegraphics[width=0.85\linewidth]{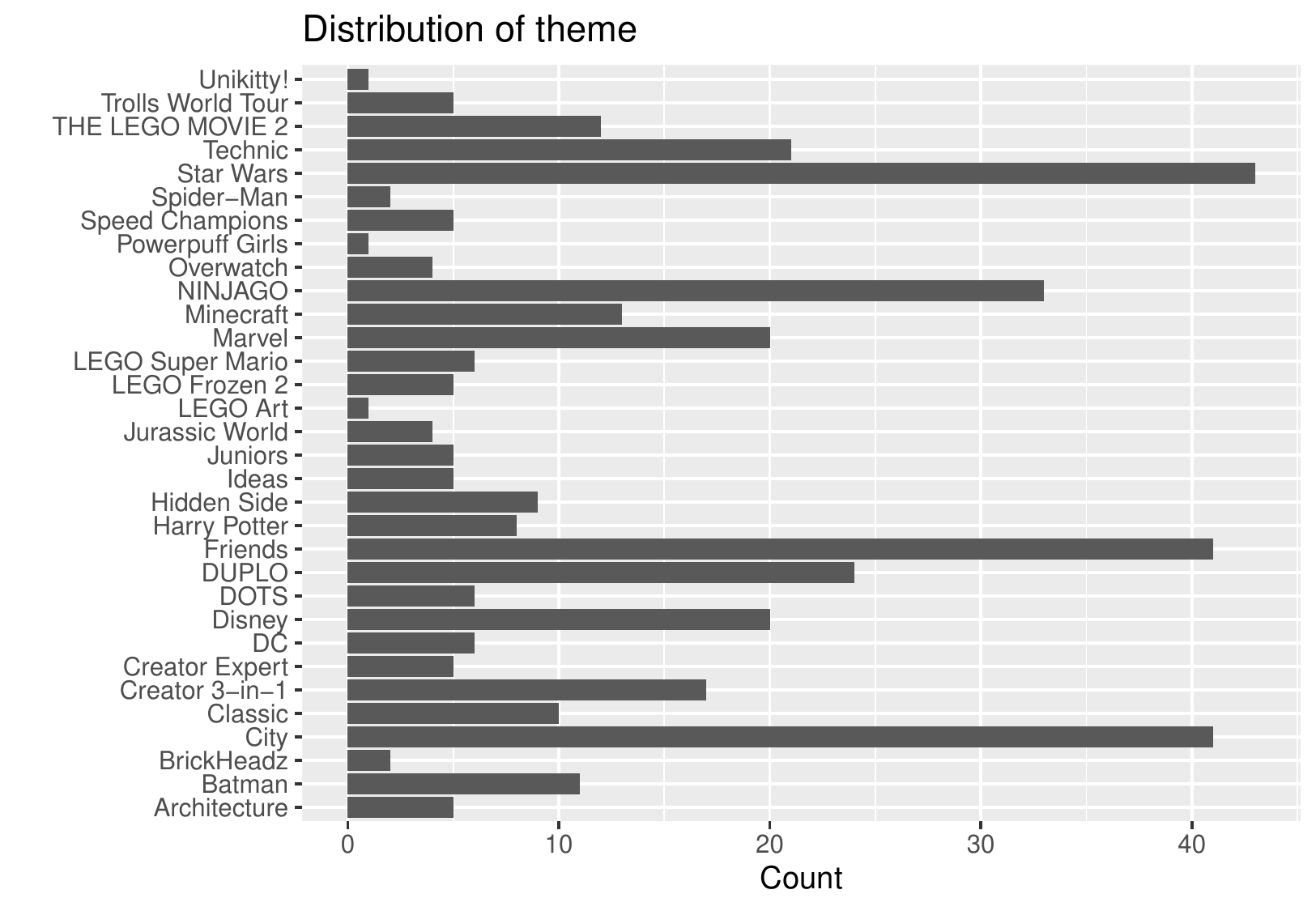} 

}

\caption{Example bar chart from LEGO in-class activity.}\label{fig:lego-orig}
\end{figure}

The LEGO brick data set and exercises modified from
\citet{peterson2021building} is used for an in-class activity about
categorical predictors and indicator variables. The data set contains
price and various characteristics of LEGO sets sold on the website
\href{https://www.brickset.com/}{brickset.com} One variable in the data
set \texttt{theme} (the theme of the LEGO set) has 32 levels, many with
very few observations. Students can make this observation when they
create a bar chart of the distribution of the variable \texttt{theme} as
shown in Figure \ref{fig:lego-orig} and discuss some of the potential
issues with including such a variable as is in a regression model.

\begin{figure}

{\centering \includegraphics[width=0.85\linewidth]{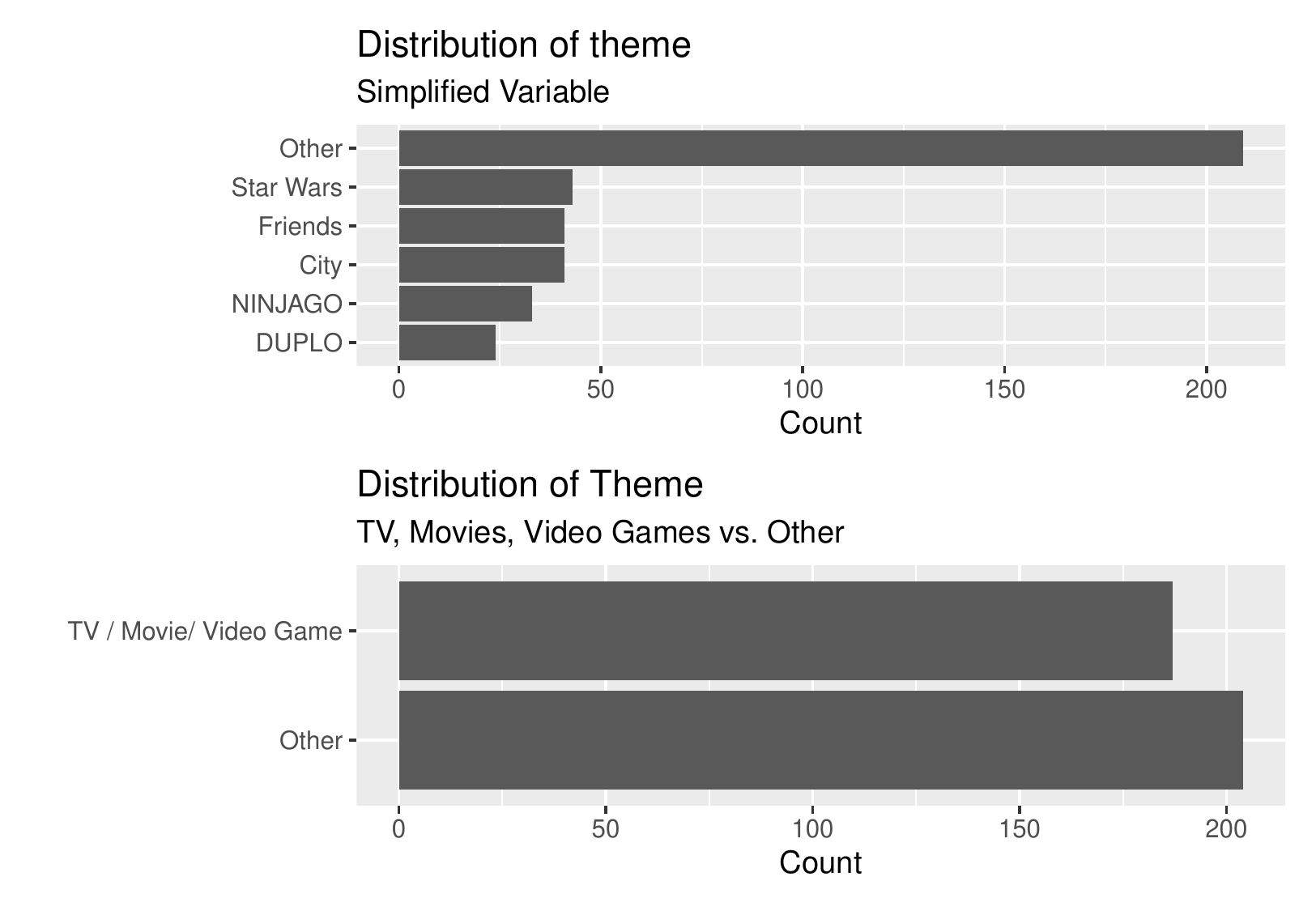} 

}

\caption{Examples of student strategies to collapse the theme variable.}\label{fig:lego-bar-collapsed}
\end{figure}

Students then consider different strategies to collapse \texttt{theme}
into a few categories that would be more feasible and informative in the
model. They consider what information they want to learn by having
\texttt{theme} in the model, how to collapse the variable to be able to
answer their analysis question, and potential advantages and limitations
to the approach. Example strategies for collapsing the variable are in
Figure \ref{fig:lego-bar-collapsed}. Once they have talked about it in
small groups, there is a whole class discussion about the advantages and
disadvantages of a few of the proposed strategies and the importance of
documenting the process of creating the collapsed variable from the
original data.

After discussing different strategies, one is chosen and that collapsed
variable \texttt{theme\_new} is used in the regression models for the
remainder of the in-class exercises. Some example questions related to
the regression models are as follows:

\begin{itemize}
\item
  \emph{Fit the model using pieces, size, and \texttt{theme\_new}, where
  \texttt{theme\_new} is the newly collapsed variable.}

  \begin{itemize}
  \tightlist
  \item
    \emph{What is the baseline level of \texttt{theme\_new}?}
  \item
    \emph{What is the interpretation of the coefficient for Star Wars?}
  \item
    \emph{What is the difference in the predicted Amazon.com price
    between a Friends set with 500 pieces and a Star Wars set with 100
    pieces, each with small pieces?}
  \end{itemize}
\end{itemize}

Another example is a homework assignment that asks students to carefully
consider the model conditions for linear regression. There are a lot of
data sets in which there is spatial dependence between observations,
thus violating the independence assumption for regression. Though
students are taught to consider potential spatial dependence, they are
often unsure about how to do so in practice. Therefore, in this homework
assignment, students consider the scenario in which they are tasked with
building a regression model that predicts the percentage of votes in a
county that were cast in-person in the 2020 election based on the
percentage of votes for Donald Trump in the 2016 election (serving as a
proxy for the political leanings of that county).

\begin{figure}

{\centering \includegraphics[width=0.85\linewidth]{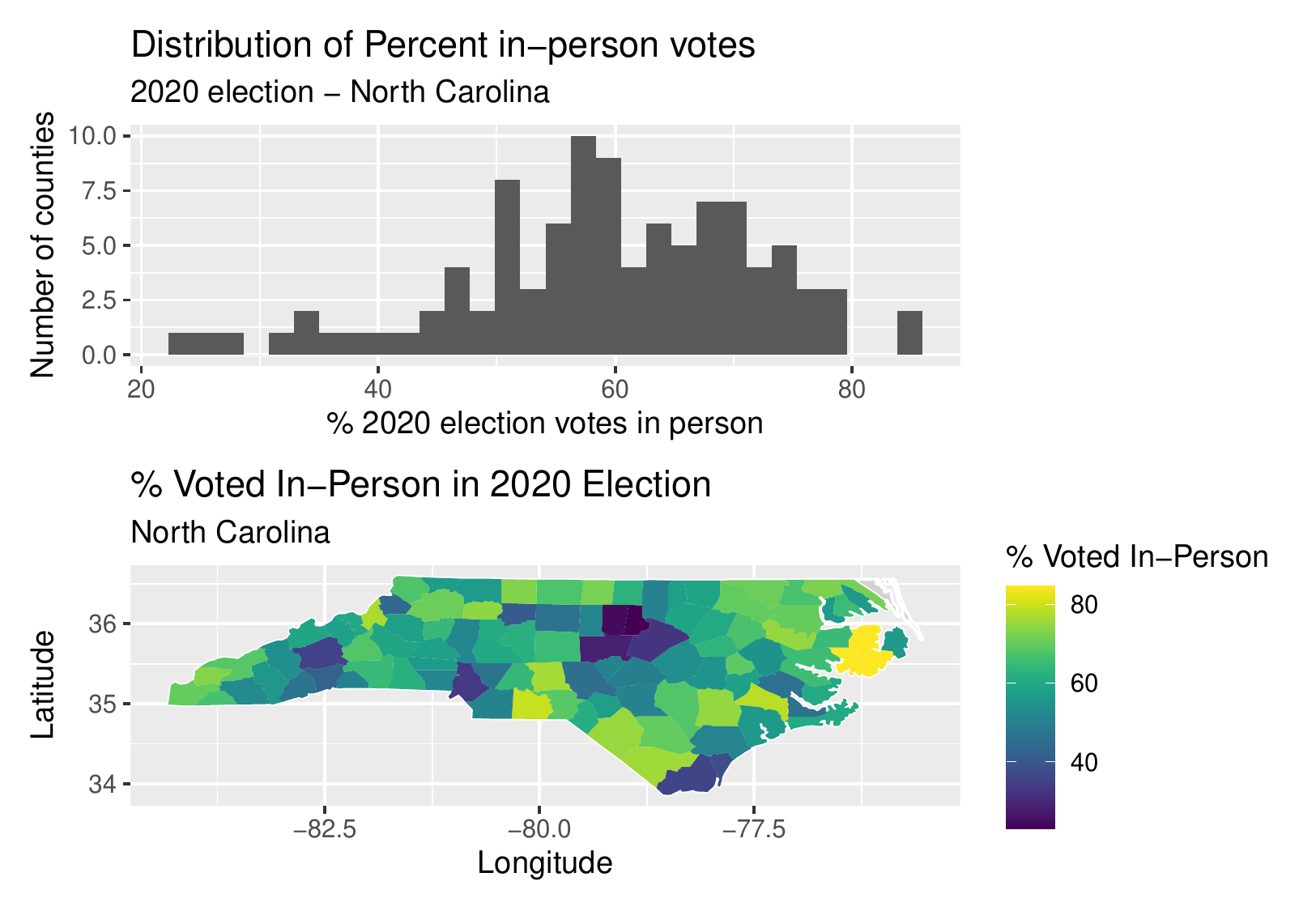} 

}

\caption{Examples of student visualizations for North Carolina voting assignment.}\label{fig:nc-data}
\end{figure}

In addition to visualizing the distribution of the response variable,
they are asked to visualize the distribution of the response variable on
a map and discuss what features are apparent in a plot such as a
histogram that are not easily apparent in the map and vice versa. They
are provided some starter code to produce the map, because spatial
visualization is not specifically taught in the course. Figure
\ref{fig:nc-data} shows examples of the visualizations students may
create.

\begin{figure}

{\centering \includegraphics[width=0.85\linewidth]{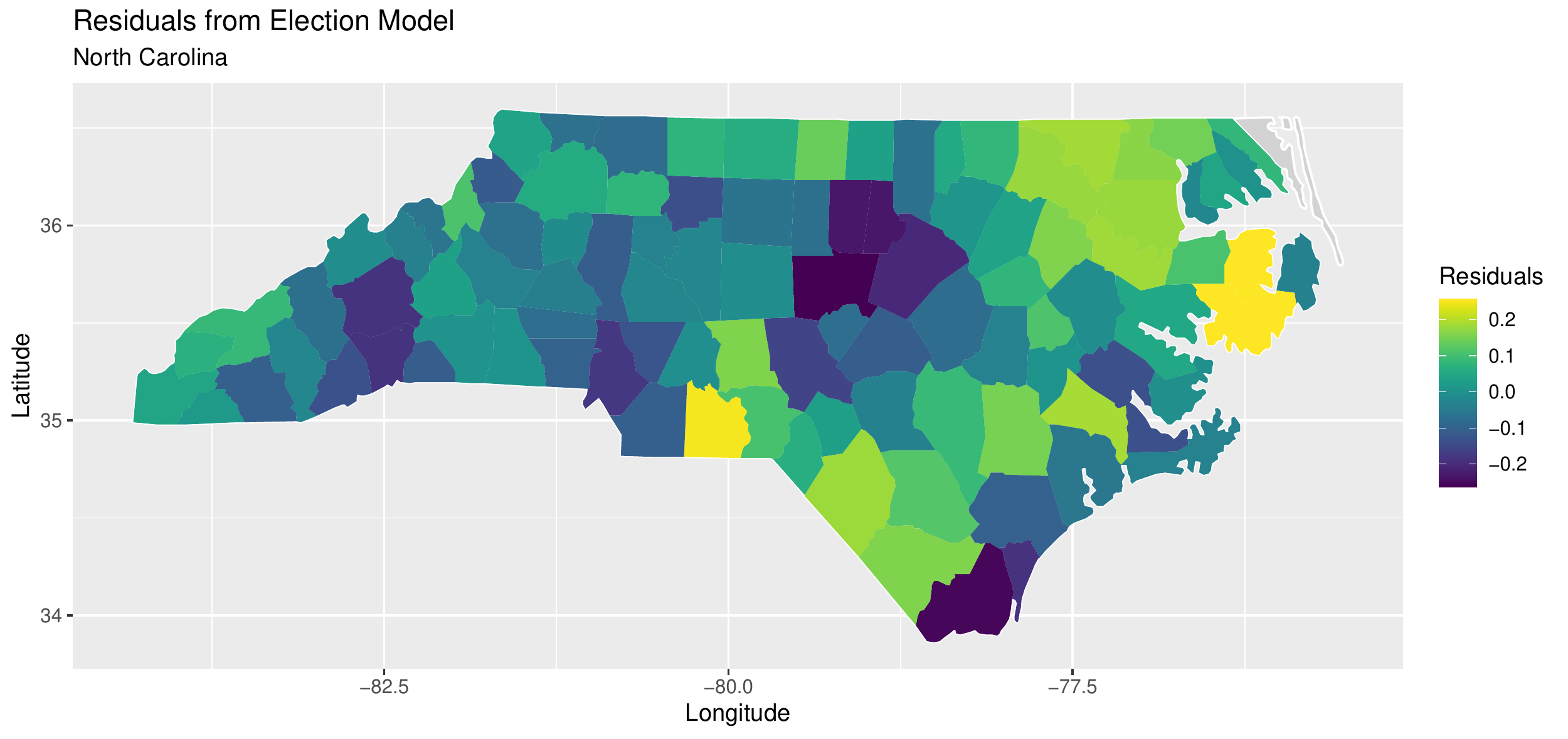} 

}

\caption{Examples of map visualizing residuals in North Carolina voting assignment.}\label{fig:nc-residuals}
\end{figure}

They are asked to fit a model and visualize the residuals on a map as
shown in Figure \ref{fig:nc-residuals}. They are then asked to assess
the independence condition using the map by answering the following
questions:

\begin{itemize}
\tightlist
\item
  \emph{Briefly explain why we may want to view the residuals on a map
  to assess independence.}
\item
  \emph{Briefly explain what pattern (if any) we would expect to observe
  on the map if the independence condition is satisfied.}
\item
  \emph{Is the independence condition satisfied? Briefly explain based
  on what you observe from the plot.}
\end{itemize}

The full in-class activity using the LEGO data and assignment using the
North Carolina county data, along with resources used to find data are
available in the supplementary materials. Additionally,
\citet{adrian2020helping} provides extensive examples on using maps to
provide context for interpretations and conclusions drawn from linear
regression.

\hypertarget{principle-2-develop-the-skills-and-computational-proficiency-for-a-reproducible-data-analysis-workflow}{%
\subsection{\texorpdfstring{Principle 2: Develop the skills and
computational proficiency for a reproducible data analysis workflow
\label{principle2}}{Principle 2: Develop the skills and computational proficiency for a reproducible data analysis workflow }}\label{principle-2-develop-the-skills-and-computational-proficiency-for-a-reproducible-data-analysis-workflow}}

Many have called for the incorporation of computing as a core part of
the statistics curriculum \citep[e.g.,][2021 \emph{Journal of Statistics
and Data Science Education} special issue ``Computing in the Statistics
and Data Science Curriculum,'' among others]{nolan2010computing}.
Developing computing skills is integrated as a key learning objective in
the course, since it is important that students have proficiency with
the computing tools that make it feasible to work with the messy
real-world data described in Section \ref{principle1}. Much more than
that, it is also in service of another one of the driving principles of
the course to equip students with the skills necessary to work with data
using a reproducible analysis workflow. While a proficiency using
statistical technology or software is required to achieve the learning
objectives in Section \ref{principle1}, these skills do not extend to
the entire data analysis workflow. Therefore in addition to proficiency
in computing, there is a more holistic approach to the data analysis
workflow, specifically writing reproducible reports using and using git
and GitHub for version control and collaboration.

There has been increased emphasis in recent years on open science and
being able to reproduce published analysis results. Using a reproducible
workflow is also good practice in industry, as it is important to write
analyses in a transparent way that can be fully replicated. By
incorporating these skills early in the statistics curriculum, students
learn practices for reproducibility as part of their analysis workflow
from the beginning, rather than having to adjust their workflow if they
learn these skills in later courses.

\hypertarget{computing-toolkit-and-infrastructure}{%
\subsubsection{Computing toolkit and
infrastructure}\label{computing-toolkit-and-infrastructure}}

The computing toolkit includes RStudio and GitHub, which were chosen for
a few key reasons. The undergraduate statistics courses are all taught
using R and many using GitHub; therefore many students will have been
introduced to R in their introductory courses, and all students will
need proficiency using these tools as they progress to the upper-level
courses. Students who have taken introductory data science have also
learned the basics of version control using Git and GitHub. Next, R and
GitHub are commonly used in academic labs and industry, so teaching with
these tools equips a broad group of students with relevant skills that
will make them more competitive as they apply for future opportunities.
Lastly, RStudio and GitHub are freely available, so all students will
have access to both platforms after they complete the course.

The computing infrastructure is modeled on the server-based system
described in \citet{ccetinkaya2018infrastructure}. RStudio is made
available through Docker containers that are set up and maintained by
the \Duke{}'s Office of Information Technology. Instructors are granted
back-end access, so they have the ability to update or install packages,
if needed. An alternative to this set up is using RStudio Cloud
(\href{https://rstudio.cloud/\%3E}{rstudio.cloud}), a cloud-based
platform set up and maintained by RStudio.

There are a few advantages to the server-based set up. First, students
access RStudio through a web browser, so they are able to use the
software from any device with access to the internet such as laptops,
tablets, or Chromebooks. The next advantage is that computing support
for students is streamlined, because all students are using the same
version of R and RStudio and have the same git configurations regardless
of computer type. Third, because git is already configured in the
RStudio containers, students are immediately able to share work between
the platforms. Lastly, the infrastructure is flexible enough to easily
accommodate in-person, hybrid, and online learning.

The main limitation to using the server-based set up is that by the end
of the course, students have only used RStudio through the web browser
rather than a version installed on their local machine. This limitation
is partially mitigated by the fact that the interface in the
university's server-based RStudio is the same as the interface in a
local installation. The largest potential hurdle for students moving
from the server to a local installation is configuring git and GitHub.
At the end of the semester, students are given instructions that direct
them to the relevant sections of the book \emph{Happy Git and GitHub for
the useR} \citep{bryan_2018_happy} for detailed step-by-step
instructions on setting up R, RStudio, and git. \Duke{}'s Office of
Information Technology provides access to a server-based RStudio for all
students at the university, so students also have the option to continue
using RStudio through the web browser.

Overall the advantages of the server-based set up outweigh the
limitations, and these advantages are not limited to RStudio. This set
up reduces many of the challenges of computing in statistics courses;
however, it is still feasible to successfully incorporate computing and
a reproducible workflow as a learning objective in a regression course
using a different computing set up and infrastructure.

\hypertarget{activities-and-assessment}{%
\subsubsection{Activities and
assessment}\label{activities-and-assessment}}

Students write responses to all in-class activities and assignments in
an R Markdown document, which facilitates writing up results in a
reproducible way. About 6 - 10\% of the total points on each assignment
(typically three to five points out of 50) are dedicated to document
formatting and implementing a reproducible workflow, which includes
using version control through regular commits and informative commit
messages. The commit history is assessed by the instructor or a graduate
teaching assistants using functions from \textbf{ghclass}, an R package
that is ``designed to enable instructors to efficiently manage their
courses on GitHub'' \citep{ghclass}. For team assignments, students are
expected to utilize GitHub for collaboration, mimicking the
collaborative workflow is commonly used in academia and industry. The
git commit history is used as part of the assessment of each students'
contribution to the team assignment, as each team member is required to
have at least one commit on these assignments.

More details about the way version control is introduced to students,
how its incorporated into assignments and assessed, and the assignment
workflow for students and the instructor in this regression course and
other courses at multiple levels of the undergraduate and graduate
curriculum are available in \citet{beckman2021implementing}.

\hypertarget{principle-3-develop-important-non-technical-skills-specifically-written-communication-and-teamwork}{%
\subsection{\texorpdfstring{Principle 3: Develop important non-technical
skills, specifically written communication and teamwork
\label{principle3}}{Principle 3: Develop important non-technical skills, specifically written communication and teamwork }}\label{principle-3-develop-important-non-technical-skills-specifically-written-communication-and-teamwork}}

The final principle for modernizing the course is the emphasis on
non-technical skills, specifically written communication and teamwork.
It is important that students are also equipped with the non-technical
skills that will make them more effective collaborators as they prepare
to conduct statistical analyses outside of the classroom.

\hypertarget{written-communication}{%
\subsubsection{Written communication}\label{written-communication}}

There has been a breadth of work on writing in undergraduate statistics,
both as a way to assess students' understanding of statistical concepts
\citep[e.g.,][]{woodard2020writing} and with the specific purpose of
developing students' communication skills
\citep[e.g.,][]{cline2008writing}. Additionally there are texts such as
the book \emph{Communicating with Data: The Art of Writing for Data
Science} \citep{nolan2021communicating} dedicated to the ``how'' of
communicating statistical results. The latter idea of effectively
communicating statistical results to a general audience is the focus of
the writing in this course.

For each assignment, points are dedicated to the quality of the
formatting and presentation of the document. ``Formatting and
presentation'' include writing all responses as a cohesive narrative in
full sentences and having a document with formatting suitable for a
professional setting. It also means using informative titles and axis
labels on graphs and neatly formatted tables and output. Students see
examples of these formatting elements in R Markdown documents for
in-class activities. Students are also provided R Markdown templates for
assignments. The templates are scaffolded, so those for early
assignments provide a lot of guidance on formatting and structuring the
responses. As the semester progresses, the scaffolding in the templates
is reduced, and by the final project students are expected to format
their document given a fairly sparse template. Because students come
from one of several introductory courses, providing the templates at the
beginning of the semester helps establish a consistent set of formatting
guidelines for the course. It also ensures students have the code
required to properly format their work without spending much time on it
in class.

In addition to formatting, points on each assignment are dedicated to
writing interpretations and conclusions that are meaningful in practice.
Therefore, students receive full credit for interpretations that are
conceptually correct and written at a meaningful scale. A rubric is used
that differentiates the accuracy of the interpretations from the
effectiveness of how the results are communicated. For example, in a
homework assignment on multiple linear regression, students are asked to
visualize and interpret interaction terms between two variables in a
data set of the price and characteristics of houses in King County,
Washington \citep{kingcounty}. They are first asked a series of
questions that assess their conceptual understanding:

\begin{itemize}
\item
  \emph{We are interested in the fitting a model using the square
  footage, whether the house has a waterfront view, and the interaction
  between the two variables to help explain variability in the price.
  Make a visualization of the price versus square footage with the
  points differentiated by waterfront. Interpret the visualization.}
\item
  \emph{Fit a model with the log-transformed price (see Lab 06 to see
  why we use log-transformed price!) as the response and sqft,
  waterfront, and their interaction as the predictors.}
\item
  \emph{Interpret the effect of square footage on the price of a house
  for}

  \begin{itemize}
  \tightlist
  \item
    \emph{houses with no waterfront view}
  \item
    \emph{houses with a waterfront view}
  \end{itemize}
\end{itemize}

Then students are asked the following prompt to assess their ability to
synthesize and communicate the results:

\emph{Use the results from the previous questions to write a short
paragraph ( \textasciitilde{} 3- 5 sentences) about the relationship
between square footage and the price of houses in King County, WA, and
how (if at all) the relationship differs based on whether the house has
a waterfront view. The paragraph should be written in a way that is
practical and can be easily understood by a general audience of home
buyers.}

In the first set of exercises, students are assessed on the accuracy of
their visualizations and interpretations. In the second exercise, the
assessment is focused on the ability to summarize the information to
derive general conclusions (not just write a list of interpretations)
and write the results using a meaningful scale. Given these criteria, a
student would not receive full credit for an technically correct
interpretation written in terms of a one square foot increase in the
size of a house, as this interpretation would not be meaningful in
practice and given the range of square feet values in the data.

In the final group project, students work in teams of three or four to
answer a research question of their choice using the methods they've
learned in the course. As part of the final project, students submit a
draft with an introduction to the data and research question, a
description of their regression modeling approach, and preliminary
conclusions. Teams peer review each other's work with a peer review
rubric that asks them to comment on specific aspects of the analysis,
including the writing and presentation. By going through the process,
students not only receive feedback on their work but they have the
opportunity to see each other's writing and garner ideas from one
another on effectively communicating results. Students also receive
feedback from a member of the teaching team at this point, either from
the instructor or a graduate teaching assistant, with some feedback
focusing on elements of the writing and presentation in addition to the
statistical results.

Students receive detailed comments along with grading based on a rubric
for the final project. Because the final project occurs at the end of
the semester, there are not opportunities for students to submitted a
revised report. One improvement to the assignment would be to introduce
the rubric for the final project early in the semester, so that students
could use it as a guide for the various writing-focused in-class
activities and assignments throughout the semester.

\hypertarget{teamwork-and-collaboration}{%
\subsubsection{Teamwork and
Collaboration}\label{teamwork-and-collaboration}}

Another primary learning objective of the course is effective teamwork
and collaboration. These are incorporated in the course through the
non-technical aspects of working with others and in technical skills
such as using GitHub for collaborative work. Given much work in academia
and industry is done on teams, it is important for students to develop
these skills in the classroom, an environment designed for learning and
growth. Working in teams also helps students learn and gain insight from
their peers. Working collaboratively ``provides students with the
opportunity to use creative problem solving and to refine leadership
skills\ldots multidisciplinary teamwork also emphasizes inclusion and
encourages diversity of thought in approaching data science problems.''
\citep{vance2021using}

Students are assigned to teams of three or four based on the results of
a ``Get to Know You'' survey at the beginning of the semester that asks
about their previous statistics and computing experience, their major or
academic interests (many students take the class before declaring a
major), and their hobbies and interests outside of the classroom. Teams
are assigned taking into account some considerations of the information
from the survey. For example, seniors and first-year students are not
put on the same team given they generally have very different
motivations for taking the course (i.e., seniors are not taking the
course to become a Statistical Science major but many first year
students may be taking the course to consider the major), and teams are
designed to reduce the possibility that a student is isolated on their
team in terms of identity and background, academic interests, or
previous experience. Students with different computing experience are
put on the same team; however, if there is too much of a disparity in
the computing experience, teams have a harder time working together, as
the more experienced student takes on most of the computational tasks.
Therefore, there is diversity in computing experience on the team to an
extent, but, for example, a student with no previous experience using R
would not be assigned to the same team as a student who rates themselves
as being a ``expert'' in R.

Barring extenuating circumstances, students work on the same team
throughout the semester on weekly computing lab assignments. By working
on the same team they have time to establish and refine their team
workflow and communication. The consistency also gives them the
opportunity to get to know each other and develop productive team
dynamics before working on the larger stakes final team project.

The first tasks for the teams are to come up with a fun team name and
fill out a team agreement that is stored in a private ``team-agreement''
GitHub repo. The primary purpose of the team agreement is to establish a
plan for communication outside of the lab sessions, including an agreed
upon method of communication and weekly meeting time outside of the lab
sessions. The work on the final project is primarily done outside of
class, so scheduling a meeting time earlier in the semester helps make
it easier to find time to collaborate on the final project. Finally,
they choose a day in which all team members will have their portion of
the assignment completed so the whole group can review the write up
before submission. Particularly with the recent challenges from remote
and hybrid collaboration, the team agreement has been helpful for
students to discuss collaboration strategies, so they are better
prepared if unexpected circumstances arise.

Students complete periodic team feedback, and the scores on feedback
account for about 2.5\% of the final course grade. The first team
feedback is only graded for completion to encourage students to provide
honest and constructive feedback without worrying about impacting their
peers' grades. It also alerts the instructor to teams that are having
trouble and need follow up. Often the issues on teams are due to gaps in
communication, therefore it is helpful to have the team agreement
available, so the team can be reminded of their collaboration plan and
discuss any necessary adjustments.

\hypertarget{discussion}{%
\section{\texorpdfstring{Discussion
\label{discussion}}{Discussion }}\label{discussion}}

\hypertarget{impact-of-this-approach}{%
\subsection{Impact of this approach}\label{impact-of-this-approach}}

Enrollment in the course has steadily increased, from 65 in Fall 2018 to
about 100 in Fall 2021. The growth has been, in part, due to the
increased popularity of the introductory courses that have encouraged
more students to continue pursuing statistics beyond the the first
course. As enrollment has increased, the course has also gained a
reputation for being one that prepares students to conduct statistical
analysis in research and internships; there are regularly requests from
students from a range of majors who want to take the course to develop
skills they'll need to conduct research in their discipline, and there
are often some requests from graduate students interested in auditing
the course to refresh their regression analysis skills. Students in the
Statistical Science major have also expressed the value of the course in
their work. Seniors in the class of 2021 were asked ``Of all the courses
you took in STA, which increased your understanding of the field the
most or provided you the best future preparation?'' Regression Analysis
was one of the top three courses mentioned in response to this question
(noting the top course being the senior capstone course).

The course has been run in flipped format described in Section
\ref{flipped-lecture} since the transition to online and hybrid formats
due to the onset of the COVID-19 pandemic. Even with the transition back
to in-person learning in Fall 2021, it is still primarily taught in a
flipped format given the pedagogical benefits. The format promotes
``productive struggle,'' the opportunity for meaningful learning as
students struggle with various concepts and ideas
\citep[\citet{lynch2018productive}]{hiebert2003developing}. Most class
time is used for hands-on exercises, so students are able to talk with
their peers and ask questions as they work through exercises. The
flipped format has also allowed for more opportunity to have in-depth
and nuanced discussions during class that encourage students to think
beyond what is ``technically correct'' to what would be useful in
practice such as those activities and exercises described in Sections
\ref{principle1} and \ref{principle2}.

Because of the emphasis on developing writing and computing skills in
the course, students have a foundation as they move into upper level
courses. Therefore instructors in the upper-level courses don't have to
spend as much time teaching students how to conduct reproducible
analyses beyond the usual content review at the beginning of the term.
Students coming in with these skills also gives instructors in
upper-level courses the opportunity to implement more of the advanced
workflow skills that students may encounter in the workplace. For
example, GitHub issues are more regularly for feedback and conversation
between the instructor and students on projects in an upper-level
undergraduate elective. Additionally, there has been more focus on the
details of writing in the elective course, such as effectively
presenting a statistical argument to various audiences and crafting a
research article, particularly when there are constraints on word count,
tables, and figures.

\hypertarget{challenges-of-implementing-these-innovations}{%
\subsection{Challenges of implementing these
innovations}\label{challenges-of-implementing-these-innovations}}

There have been some challenges to incorporating these innovations in
the course. The first is identifying interesting and complex data sets
that are still accessible to new learners. Having to address model
conditions can provide a learning experience and motivate concepts such
as log transformations on the response and predictor variables. It can
also be an opportunity to get students excited about later units in the
course or future courses where they will learn how to deal with data
that do not neatly fit the conditions for linear regression. For
example, students see how log transformations can be used to model
non-linear relationships by examining the relationship between movie
budgets and revenues based data from
\href{https://www.imdb.com}{IDMB.com} in the first week of class. There
is not a lot of time spent on log transformation at the moment, but it
gives students a glimpse of the type of data they'll be able to analyze
using techniques from the course.

As the course progresses, examples such these can also be used to
facilitate more nuanced discussions about statistical inference and
model conditions, such as different approaches (simulation-based
vs.~mathematical models) and the robustness of these approaches to
violations in conditions. These discussions expose students to a variety
of realistic modeling scenarios and gives them practice developing the
judgment and decision-making skills needed to use effectively use
modeling in practice. It is important, however, that students first have
an understanding about what model conditions are and how they relate to
inference for regression, so they understand the motivation for checking
conditions even if inferential conclusions can be drawn from models with
some violations.

The next challenge is having the ability to provide meaningful
individual and iterative feedback to each student on writing exercises,
particularly long-form writing where students must synthesize
information from a full analysis to draw conclusions. As the course has
increased in size, it has become not feasible to provide such detailed
feedback to individuals regularly during the semester. There are some
homework assignments in which students are asked to summarize results in
one to two paragraphs as shown in Section \ref{principle3} or where they
are asked to conduct an abbreviated data analysis; however, the only
assignment where students get iterative feedback on drafts of their
writing with the opportunity to improve and resubmit is in the final
group project.

The teaching team consists of undergraduate and graduate teaching
assistants who help with grading assignments. It can be challenging to
train teaching assistants, especially undergraduates who themselves are
still relatively new learners, to accurately and effectively grade
open-ended response. New learners can still have difficulty
differentiating responses that are incorrect versus those that are still
correct but presented differently than what is in the
instructor-provided solutions manual. This has been partially mitigated
by using Gradescope, an online grading rubric platform
\citep{gradescope}, and creating specific rubric items to identify key
parts of an interpretation or analysis; however, without additional
feedback to provide context to the rubric items, the very specific point
system can seem more punitive to students rather than an indication of
misunderstanding important concepts.

\hypertarget{whats-next-for-the-course-and-curriculum}{%
\subsection{What's next for the course and
curriculum}\label{whats-next-for-the-course-and-curriculum}}

In the most recent semester, innovations in the course have been more
focused on modernizing the choice of topics and methods in the course.
The most substantial change has been an emphasis on using regression
models for prediction in addition to inference. Topics such as
cross-validation were introduced early and carried as a theme throughout
the semester. Other modern methods such as simulation-based inference
were taught as core content in the course. The course also incorporated
the use of \textbf{tidymodels}
(\href{https://www.tidymodels.org/}{tidymodels.org}), a suite of
modeling packages that aim to provide a more unified syntax and
framework for modeling in R and more easily facilitate a regression
workflow for both inference and prediction. The new content has created
opportunities for more discussions about how the primary analysis
objective (inference or prediction) is used to help inform the
methodology, along with how to assess model conditions, model
diagnostics, and their relative importance for different methods. To
accommodate the addition of these topics there is less time in the
course spent on ANOVA and simple linear regression, though a large
portion of the course is still dedicated to multiple linear regression.

There is also ongoing work with faculty in \Duke{}'s writing center to
develop more writing exercises that are feasible for the large classes,
so students receive more individual and iterative feedback on writing
throughout the semester.

\hypertarget{conclusion}{%
\subsection{Conclusion}\label{conclusion}}

As the ability to analyze data has become an increasingly important
skill in academia, industry, and every day life, it is important that
the statistics curriculum is designed to prepare students to responsibly
work with modern data outside of the classroom. While there has been a
lot of work on this front in the modernization of introductory
statistics courses and the development of introductory data science
courses, it is critical that the innovation extends beyond the
introductory course. Three principles used to modernize an undergraduate
regression analysis course, the second statistics course for many
students, were presented. After implementing these innovations in the
first and second courses, there has been increased student interest in
the discipline and more opportunity to continue developing these skills
in subsequent courses. While there are many benefits to this approach,
instructors should keep in mind some of the cautions presented earlier,
such as not discouraging students by consistently using data that
requires methods beyond the scope of the course, being mindful of the
student and instructor workload while at the same time finding
opportunities for meaningful feedback, particularly on open-ended
written work.

Though these principles were designed with the second course in mind,
they are applicable and can benefit students throughout the curriculum.
By incorporating a unified set of principles that influence the design
of courses throughout the curriculum, students will have a more
streamlined experience where they can continue developing their
computing and communications alongside their statistical knowledge and
literacy.

\hypertarget{acknowledgements}{%
\section*{Acknowledgements}\label{acknowledgements}}
\addcontentsline{toc}{section}{Acknowledgements}

I thank the editor, associate editor, and anonymous reviewers for their
thoughtful comments and suggestions. I also thank Beth Chance for her
mentorship and thoughtful feedback throughout the process of writing
this manuscript.

\hypertarget{supplementary-materials}{%
\section*{\texorpdfstring{Supplementary materials
\label{supplement}}{Supplementary materials }}\label{supplementary-materials}}
\addcontentsline{toc}{section}{Supplementary materials
\label{supplement}}

Supplementary materials for the article including links to course
websites, sample activities and assignments, and other resources are
available at
\href{https://github.com/matackett/modernize-regression}{github.com/matackett/modernize-regression}.

\hypertarget{data-availability-statement}{%
\section*{Data Availability
Statement}\label{data-availability-statement}}
\addcontentsline{toc}{section}{Data Availability Statement}

No new data were created or analyzed for this manuscript. Data used in
the example in-class activity and assignment are available in the
supplementary materials.

\bibliographystyle{agsm}
\bibliography{modernreg-revised.bib}

\end{document}